# Excluded Volume Effects in Gene Stretching

Pui-Man Lam
Physics Department, Southern University
Baton Rouge, Louisiana 70813

Abstract

We investigate the effects excluded volume on the stretching of a single DNA in solution. We find that for small force F, the extension h is not linear in F but proportion to $F^{\gamma}$, with $\gamma=(1-\nu)/\nu$, where $\nu$ is the well-known universal correlation length exponent. A freely joint chain model with the segment length chosen to reproduce the small extension behavior gives excellent fit to the experimental data of $\lambda$-Phage DNA over the whole experimental range. We show that excluded volume effects are stronger in two dimensions and also derive results in two dimensions which are different from the three dimensional results. This suggests experiments to be performed in these lower dimensions.

I.  INTRODUCTION

Many forms of polymers or macromolecules such as proteins, lipids, fatty acids and DNA, are vital for life. The DNA is a very large molecule. For instance the DNA in one human cell can encode approximately 100,000 genes. It is partitioned into 46 chromosomes, 23 from each parent and is normally twisted and folded into the nucleus of a cell a few micrometers in width. In solution the human DNA consists of two strands of polymerized nucleotides, twisted into a right hand helix. New technologies makes it possible to study single molecules of DNA. Such studies can give insights for development of laboratory techniques for analyzing, fractionating and sequencing DNA. One such study investigates how a single DNA stretches when it is pulled with a force at its two ends, using micromanipulation tools such as optical tweezers or microneedles to grasp the molecule by its two ends. Stretching the DNA can provide a much clearer picture under the microscope. As the average end-to-end separation h increases, a force arises that opposes the stretching. This force is entropic in nature since the number of configurations of the polymer decreases as h increases, leading to a decrease in the entropy. Such a force versus extension curve $F(h_z)$ for a force applied in the z-direction had been directly measured for a 97-kb λ-Phage DNA dimer [1]. The experimental data can be fitted remarkably well using a worm-like chain (WLC) polymer model [2]. However, in the WLC model, excluded volume effect was not taken into account. For a polymer in solution, excluded volume effect is always present, except perhaps at the tricritical θ-point. Therefore there is in general no reason for this effect to be negligible for DNA in solution. In this paper we investigate the effect of excluded volume in DNA stretching. We find that for a week force F applied at the ends of the molecule, the extension $h_z$ in the direction of the force is not linear in F, but rather proportional to $F^\gamma$, with $\gamma=(1-\nu)/\nu$, where ν is the well-known universal correlation length exponent. Using the result ν=0.6, [3] this gives γ=2/3. We find that the experimental data for λ-Phage DNA at small F is actually in agreement with this behavior with γ≈0.64. Including the excluded volume effect in a freely joint chain (FJC) model by choosing the segment length to reproduce the small extension behavior gives excellent fit to the full range of experimental data.

In two dimensions, the exponent ν is exactly known to be ¾[3]. This will give an extension proportional to $F^{1/3}$ in a week force F. In the FJC model, the functional form of the force versus extension curve $F(h_z)$ is different from that of the three dimensional case. Again the excluded volume effects can be included by choosing the segment length in the FJC to reproduce the small extension behavior. This suggests experiments to be performed in two-dimensional geometry to test the stronger effect of excluded volume in lower dimensions.

In section II we will review the models in which excluded volume effect is neglected, including the Gaussian chain and the FJC model. In section III we will discuss DNA model with the excluded volume effect and use it to fit the experimental data. In section IV we will discuss the two dimensional case. Section V is the conclusion.

## II. MODELS WITHOUT EXCLUDED VOLUME EFFECT

### A. GAUSSIAN CHAIN

In the Gaussian chain model, a long DNA molecule at equilibrium in solution has random walk statistics. The probability distribution p(h) of its end-to-end distance h is approximately Gaussian [4]:

$$p(h) = \left(\frac{3}{2\pi <R>^2}\right)^{3/2} \exp\left(-\frac{3h^2}{2<R>^2}\right) \qquad (1)$$

where <R> is the average end-to-end distance. For fixed h, the entropy of the DNA is $S=k_B \log[p(h)]$, where $k_B$ is the Boltzmann's constant. For a fixed end-to-end distance, the force in the z-direction is given by

$$F_z = -T\frac{\partial S}{\partial h_z} = 3k_B T h_z / <R>^2 \qquad (2)$$

For a polymer of with a contour length L, the persistence length P can be defined in the Gaussian chain as P=<R>$^2$/(2L) [4]. This persistence length is independent of the contour length since for a Gaussian chain, <R>$^2$ ~L. In terms of the persistence length P, equation (2) can be written as

$$h_z = \frac{2LP}{3k_B T} F_z. \qquad (3)$$

The extension is linear in the force, so the DNA behaves as a Hookean spring with zero natural length and a temperature dependent effective spring constant.

B. FREELY JOINT CHAIN

In the freely joint model [5], the chain of contour length L consists of (L/b) freely joint segments each of length b. When this chain is stretched at its two ends with a force F, each segment is pulled at both ends with this force F and thus tends to align it in the direction of F, while thermal fluctuations tend to orient it in random directions.

For a force in the z direction and denoting the angle between $F_z$ and a particular segment as θ, the partition function of the chain can be written as

$$Z = \left[\int_0^{2\pi} \exp\left(\frac{bF_z \cos(\theta)}{k_B T}\right) 2\pi \sin(\theta)d\theta\right]^{L/b} = \left[\frac{4\pi k_B T}{bF_z}\sinh\left(\frac{bF_z}{k_B T}\right)\right]^{L/b} \qquad (4)$$

The average extension of the chain in the z direction is given by

$$h_z = k_B T \frac{\partial}{\partial F}\log Z = L\left[\coth\left(\frac{bF}{k_B T}\right) - \frac{k_B T}{bF}\right] \equiv L\Lambda\left(\frac{bF}{k_B T}\right) \qquad (5)$$

where $\Lambda(x)=\coth(x)-x^{-1}$ is the Langevin function. For small x, $\Lambda(x) \approx x/3$. Therefore for small F, $h_z \approx LbF/(3k_BT)$. Comparison with equation (3) gives b=2P. Substituting into equation (5) then gives

$$h_z = L\Lambda\left(\frac{2FP}{k_BT}\right) \tag{6}$$

II.  MODELS WITH EXCLUDED VOLUME EFFECT

For a polymer with excluded volume effect, the probability distribution of the end-to-end distance is given by [6]

$$p(h) = A\left(\frac{h}{<R>}\right)^a \exp\left[-\frac{3}{2}\left(-\frac{h}{<R>}\right)^d\right], \tag{7}$$

where A is a normalization constant. The exponent $\delta$ is related to the universal exponent for the average end-to-end distance by $\delta=1/(1-\nu)$ [7]. The exponent $\alpha$ is strictly zero at the upper critical dimension $d_c=4$, when $\nu=1/2$ and the probability distribution reduces to the form given in equation (1). We will show in the following that $\alpha$ should be zero in all dimensions.

The entropy is given by

$$S = k_BT \log p(h) = k_B\left[\log A + a\log\frac{h}{<R>} - \frac{3}{2}\left(\frac{h}{<R>}\right)^d\right] \tag{8}$$

The force exerted at the ends of the chain is given by

$$F = -T\frac{\partial S}{\partial h} = -k_BT\left[a\frac{<R>}{h} - \frac{3d}{2<R>^d}h^{d-1}\right] \tag{9}$$

Unless α=0, the first term would imply the unphysical result that the force F becomes infinite as the extension h goes to zero. Putting α=0 in equation (7), the component of the force in the z direction is given by

$$F_z = -T \frac{\partial S}{\partial h_z} = \frac{3dk_B T}{2<R>^d} h_z (h^2)^{(d-2)/2} \qquad (10)$$

The average force $F_z$ can be obtained by taking $h_x^2 = h_y^2 = h_z^2$:

$$F_z = \frac{3dk_B T}{2<R>^d} h_z (3h_z^2)^{(d-2)/2} = \frac{3^{d/2} dk_B T}{2<R>^d} h_z^{d-1} \qquad (11)$$

The extension $h_z$ of the chain in the z direction due to a force $F_z$ in the z direction is given by

$$h_z = \left[ \frac{2F<R>^d}{3^{d/2} dk_B T} \right]^{\frac{1}{d-1}} \qquad (12)$$

Therefore with excluded volume effect, the extension is not linear in F for δ≠2.

In Figure 1 we show the double logarithmic plot of the experimental data for the extension versus force of a 97-kb λ-Phage DNA dimer [1]. The slope of the curve at small extension is about 0.64. This is in good agreement with the exponent 1/(δ-1)=2/3 obtained using the known value of ν=0.6 [3]. This shows that excluded volume effect is important for DNA.

In the freely joint chain model we again assume a chain consisting of L/b freely joint segments each of length b. Then the average extension in the direction of the force is again given by equation (5). The effects of excluded volume can be included in the FJC model by choosing the segment length b to reproduce the small extension behavior. Comparing the small extension result of that equation $h_z = LbF/(3k_BT)$ with equation (12), we obtain for the segment length

$$b = \frac{3k_BT}{FL}\left[\frac{2F<R>^d}{3^{d/2}dk_BT}\right]^{\frac{1}{d-1}} \tag{13}$$

Substituting this into equation (5) we obtain the expression for the extension in the direction of the force for a chain with excluded volume effect

$$h_z = L\Lambda\left(\frac{3}{L}\left[\frac{2F<R>^d}{3^{d/2}dk_BT}\right]^{\frac{1}{d-1}}\right) \tag{14}$$

In contrast to the case of a Gaussian chain, the persistence length defined by $P=<R>^2/(2L)$ is not a constant but would depend on the contour length L. Therefore we use here the average end-to-end distance $<R>$ as a parameter, rather than the persistence length. We have fitted the experimental data [1] to equation (14), with $\delta=1/(1-\nu)=2.5$, but varying the parameters L and $<R>$ to minimize the least square deviation. The best fit is obtained with $<R>=3.19\mu m$ and $L=32.8\mu m$. In Figure 2 we show the result of our fit with the experimental data of the extension versus force. The dashed lines are the results of the freely joint chain with no excluded volume effect. The value P=31nm corresponds to best least square fit with the data and P=50nm corresponds to a fit to the small extension data. We see that the fit to the experimental data is excellent for the model with excluded volume effect. A least square fit varying all three parameters $<R>$, L and $\delta$ yields $\delta=2.6$. This shows that the form $\delta=1/(1-\nu)$ is in fact very good.

We have also fitted the more recent data of Wang et al [8] to equation (14). Their experimental data were obtained using a double stranded DNA of much shorter size (~1µm). Wang et al. found that their data could be fitted well with the theoretical forms of Marko and Siggia [9], (with persistence length P=43.5nm, contour length L=1.31µm) and Odjik [10] (with persistence length P=41.2nm, contour length L=1.32µm). Both forms take into account also elasticity theory. However, they found that their experimental data were not well fit by the freely joint chain expression over any force region. This finding in fact supports conclusions of previous work with double-stranded DNA molecules. The freely joint chain model, however can be successfully applied to describe singly stranded DNA. Since our theory with excluded volume is based on the freely joint chain model we do not expect it to fit well the data of a double stranded DNA. In spite of this we show in Figure 3 our best fit to the data of Wang et al. [8]. The dashed curve is obtained using the freely joint chain model without excluded volume, with persistence length P=14nm and contour length L=1.3µm. The solid curve is the best fit using our expression (14) with $<R>$=0.394µm, L=1.3µm and $\delta$=2.5. Even though both the freely joint chain and our equation (14) do not fit the data well we find that our expression seem to fit better over most of the data range. We also find that the largest slope in the data is about 0.25. Therefore even though the force-extension relation is non-linear, the exponent is much smaller than the value 2/3 predicted by our theory. This is again due to the double stranded nature of the DNA.

It should be pointed out that our expression (14) applies only when the chain is not yet close to fully extended. When the chain is close to fully extended, the freely joint chain expression (6) should be better, because then the excluded volume effect must be negligible. One could, in principle think of a way of bridging the two expressions (6) and (14). However, it is not known at what extension, or force, is the excluded volume effect in the chain negligible. Only when the chain is fully extended it is clear that the excluded volume effect is negligible. But the fully extended configuration is only one among a large number of configurations whose number increases exponentially with the length of the chain. For a chain that is close to but not yet fully extended the critical force for the crossover can only be determined by comparing the fit of the experimental data to (6) and (14) to see at what extension is the expression (6) a better fit than (14). Looking at Figure 2, one can see that the fits to the data from (6) and (14) are almost indistinguishable. Therefore we prefer to leave (14) as it is, keeping this condition in mind.

III. DNA IN TWO DIMENSIONS

A. WITHOUT EXCLUDED VOLUME EFFECT

The partition function for a freely joint chain given in equation (4) is valid only for three dimensions. In two dimensions it has to be replaced by

$$Z = \left[\int_0^{2\pi} d\theta \exp\left(\frac{Fb\cos\theta}{k_B T}\right)\right]^{L/b} = \left[2\pi I_0\left(\frac{Fb}{k_B T}\right)\right]^{L/b} \quad (15)$$

where $I_0$ is the modified Bessel function of order zero. The average extension in the direction of the force is given by

$$h_z = k_B T \frac{\partial}{\partial F} \log Z = L I_1\left(\frac{Fb}{k_B T}\right) \bigg/ I_0\left(\frac{Fb}{k_B T}\right) \quad (16)$$

where $I_1$ is the modified Bessel function of order one. From the behaviors of $I_0$ and $I_1$ for small arguments one obtains

$$h_z = \frac{LFb}{2k_BT}, F \to 0 \qquad (17)$$

Comparing this to the Gaussian chain result for the two dimensional case, $h_z=PLF_z/(k_BT)$, we find b=2P. Substituting this into equation (16), we find

$$h_z = L I_1\left(\frac{Fb}{k_BT}\right) \bigg/ I_0\left(\frac{Fb}{k_BT}\right) \qquad (18)$$

for the extension of a two-dimensional DNA in direction of the force F.

B. WITH EXCLUDED VOLUME EFFECT

To take into account excluded volume effect we use equation (10) with $h^2=h_x^2+h_z^2=2h_z^2$ and find

$$F_z = \frac{dk_BT}{<R>^d} h_z (2h_z^2)^{\frac{d-2}{2}} = 2^{\frac{d-2}{2}} \frac{dk_BT}{<R>^d} h_z^{d-1} \qquad (19)$$

From this we find

$$h_z = \left[\frac{2^{(2-d)/2} F <R>^d}{dk_BT}\right]^{\frac{1}{d-1}} \qquad (20)$$

In a freely joint chain model with (L/b) segments, each of length b, the extension in the direction of the force is given by (16), with the small F behavior given by equation (17). Again the excluded volume effect can be included by choosing the segment length in the FJC model to reproduce the small extension behavior. From equations (17) and

(20) we find

$$b = \frac{2k_B T}{LF} \left[ \frac{2^{(2-d)/2} F <R>^d}{d k_B T} \right]^{\frac{1}{d-1}} \quad (21)$$

Substituting this value of b into equation (16) we have

$$h_z = LI_1\left(\frac{2}{L}\left[\frac{2^{(2-d)/2} F <R>^d}{d k_B T}\right]^{\frac{1}{d-1}}\right) \bigg/ I_0\left(\frac{2}{L}\left[\frac{2^{(2-d)/2} F <R>^d}{d k_B T}\right]^{\frac{1}{d-1}}\right) \quad (22)$$

In two dimensions, the exponent for the end-to-end distance is known exactly to be $\nu=3/4$ [3]. This gives $\delta=1/(1-\nu)=4$, a much larger deviation from the value $\delta=2$ in the case of no excluded volume effect. Only two parameters L and <R> in equation (22) are left to fit experimental data. However since no experimental data is known for DNA in two dimensions, we will use the same values of L and <R> obtained in three dimensions. The results are shown in Figure 4 for both with and without excluded volume effect. We suggest experiments to be performed in two dimensions to test the stronger effects of excluded volume in lower dimensions.

IV.   CONCLUSION

We investigate the effects excluded volume on the stretching of a single DNA in solution. We find that for small force F, the extension h is not linear in F but proportion to $F^\gamma$, with $\gamma=(1-\nu)/\nu$, where $\nu$ is the well-known universal correlation length exponent. A freely joint chain model with the segment length chosen to reproduce the small extension behavior gives excellent fit to the experimental data of $\lambda$-Phage DNA over the whole experimental range. The parameters used in the fitting are the contour length L and the average end-to-end distance <R>. The best fit is obtained with <R>=3.19µm and L=32.8µm. The fit to the experimental data is just as good as the worm like chain model [2], in which only the elastic energy is taken into account by treating the chain as a uniform elastic rod. It is possible that both effects are present in DNA. We show that excluded volume effects are stronger in two dimensions and also derive results in two dimensions which are different from the three dimensional results. This suggests experiments to be performed in these lower dimensions

**Acknowledgement:** This work has been supported by the Department of Energy grant DE-FG02-97ER25343

**FIGURE CAPTIONS**

Figure 1: Double logarithmic plot of extension versus force for λ-phage DNA. The slope of the curve at small extension is about 0.64

Figure 2: Extension versus force. The solid dots are experimental data of λ-phage DNA. The solid line is for freely-joint chain with excluded volume effects, using the best least square fit values for <R> and L and fixed value of δ=2.5. The dashed curves are results of freely joint chain, for two values of the persistence length.

Figure 3: Extension versus force. The solid dots are experimental data of ref. [8]. The solid line is for free-joint chain with excluded volume effects, using the best least square fit values of <R> and L and fixed value of δ=2.5. The dashed curve is the result of the freely joint chain.

Figure 4: Extension versus force for a two dimensional polymer in the freely joint chain model. The dashed curve is for the freely joint chain model in two dimensions without excluded volume, with persistence length P=50 nm. The dotted curve is the same thing for three dimensions, for comparison. The solid curve is for the freely joint chain model in two dimensions with excluded volume, using parameter values of <R> and L obtained in three dimensions.

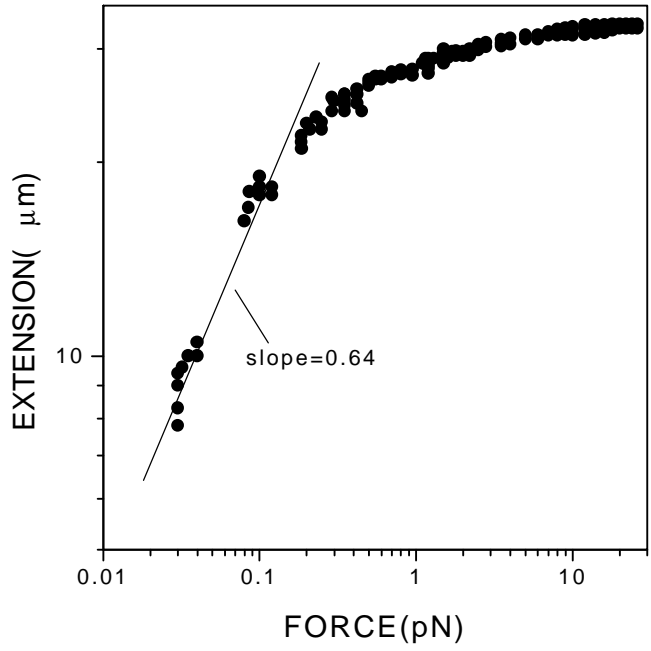

Figure 1

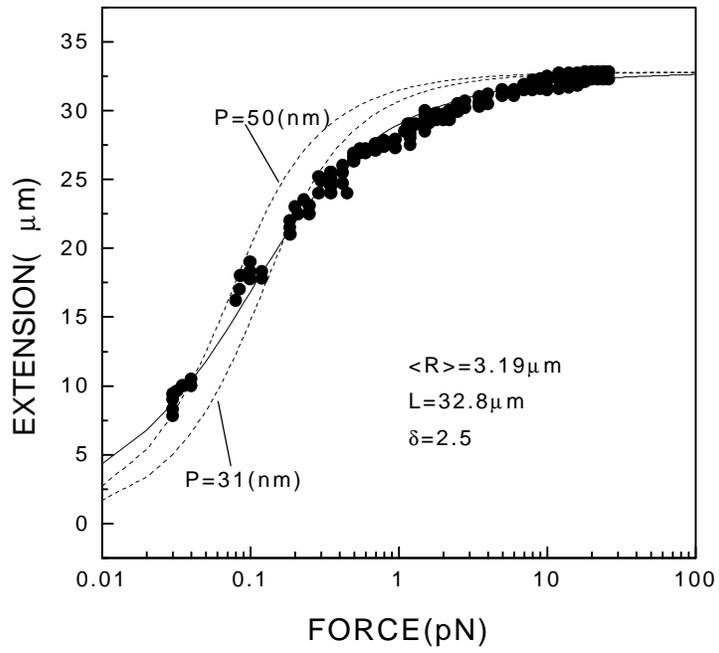

Figure 2

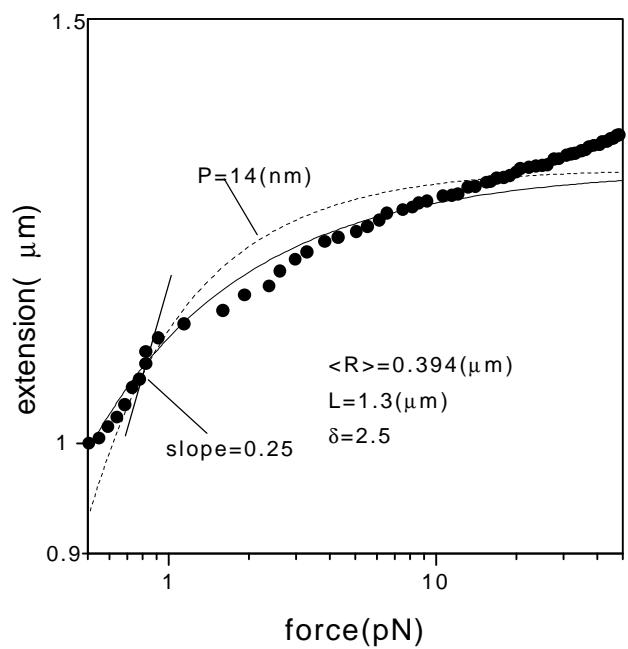

Figure 3

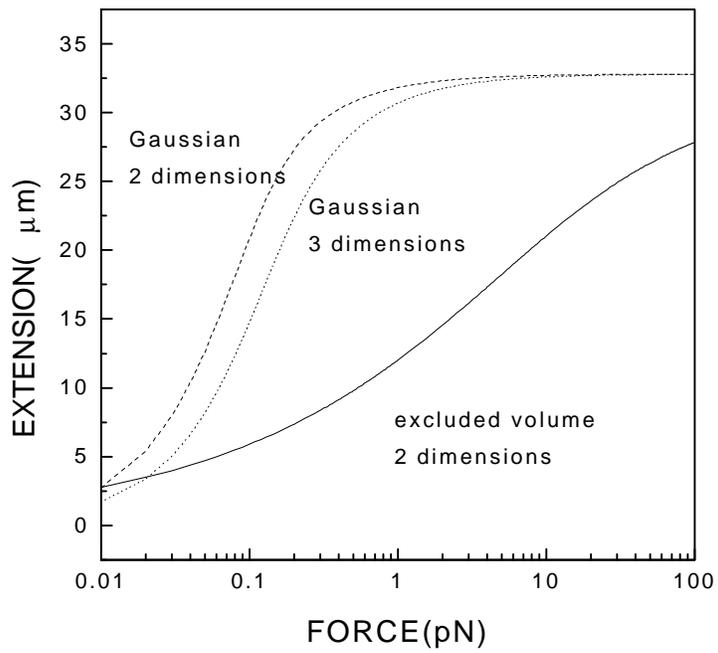

Figure 4